\documentclass[12pt]{article}%
\usepackage{amsmath}%
\usepackage{amsfonts}%
\usepackage{amssymb}%
\usepackage{graphicx}

\begin{document}

\title{Biaxial nematic phase in the Maier-Saupe model for a mixture of discs and
cylinders }
\author{E. F. Henriques and S. R. Salinas\\Instituto de F\'{\i}sica, Universidade de S\~{a}o Paulo\\Caixa Postal 66318, CEP 05314-970, S\~{a}o Paulo, Brazil}
\date{14 October 2011}
\maketitle

\begin{abstract}
We analyze the global phase diagram of a Maier-Saupe lattice model with the
inclusion of disorder degrees of freedom to mimic a mixture of oblate and
prolate molecules (discs and cylinders). In the neighborhood of a Landau
multicritical point, solutions of the statistical problem can be written as a
Landau-de Gennes expansion for the free energy. If the disorder degrees of
freedom are quenched, we confirm the existence of a biaxial nematic strucure.
If orientational and disorder degrees of freedom are allowed to thermalize,
this biaxial solution becomes thermodynamically unstable. Also, we use a
two-temperature formalism to mimic the presence of two distinct relaxation
times, and show that a slight departure from complete thermalization is enough
to stabilize a biaxial nematic phase.

\end{abstract}

\section{Introduction}

The transition between a uniaxial nematic structure and an orientationally
disordered phase is perhaps the most investigated and best characterized phase
transition phenomenon in liquid crystalline systems \cite{deGennes}. This weak
first-order transition is quite well described by the mean-field theory of
Maier and Saupe \cite{Gramsbergen}\cite{Chandrasekhar}\cite{Singh}, which can
also be formulated in terms of a fully-connected statistical lattice
Hamiltonian \cite{Figueiredo}. The existence of a biaxial nematic nematic
phase, however, and the transitions between different types of nematic
structures, which have been proposed on the basis of phenomenological
calculations for systems with intrinsically biaxial molecular groups
\cite{Freiser}, turned out to be much more difficult to characterize
experimentally \cite{Luckhurst}. Although there have been some recent reports
of a biaxial nematic structure in thermotropic liquid crystalline systems
formed by bent-core or boomerang-shaped molecules \cite{Madsen}, a biaxial
phase has been first characterized in the phase diagram of a lyotropic liquid
mixture \cite{YuSaupe}\cite{Galerne}, which is better represented by a lattice
model of shape-disordered uniaxial molecules. We then revisit the problem of a
Maier-Saupe lattice model, with the inclusion of extra degrees of freedom to
mimic a mixture of oblate and prolate molecules (discs and cylinders).
Slightly different versions of this problem have been analyzed by different
authors , with some conflicting results \cite{Palffy, Henriques, Photinos,
Berardi}. According to the conclusions of a recent calculation for a
shape-disordered Maier-Saupe model with restricted orientations \cite{Carmo1},
we point out that the introduction of two sets of degrees of freedom opens the
possibility of choosing different relaxation times, with different outcomes
for the thermodynamic stability of a biaxial nematic structure.

We then formulate and analyze a Maier-Saupe lattice model for a mixture of
discs and cylinders. This problem includes orientational (quadrupolar) and
shape-disordered (discs and cylinders) degrees of freedom, which might be
associated with different relaxation times. Solutions can be obtained by the
application of well-known methods of statistical mechanics. First, we treat
the case of fixed (frozen) disorder, as in a typical problem of a disordered
solid state system. In this quenched case, disorder (shape) degrees of freedom
are fixed, frozen, while the orientational degrees of freedom are allowed to
thermalize during experimental times. We then treat the case of annealed
(thermalized) disorder, in which both orientational and disorder degrees of
freedom are allowed to reach thermodynamic equilibrium during experimental
times, and which is certainly more adequate to explain the behavior of a
liquid system. In agreement with previous calculations for similar models, in
the quenched case we show the existence of a biaxial nematic structure for
typical distributions of shape-disorder \cite{Henriques}. In the thermalized
case, however, there is a biaxial nematic solution of the model equations, but
it turns out to be thermodynamically unstable (in agreement with older
calculations by Palffy-Murhoray and collaborators \cite{Palffy}). We then
resort to a formalism based on two heat reservoirs, at distinct temperatures,
which is conceived to describe an intermediate situation, between fully fixed
and fully thermalized distributions of disorder variables \cite{Nieuwenhuizen}%
. As in the calculations for the Maier-Saupe model with restricted
orientations \cite{Carmo1}, we show that a small difference of temperatures,
which is equivalent to a slight departure from complete thermalization, is
already sufficient to produce a stable biaxial nematic phase.

It should be pointed out that we write closed-form solutions of the
statistical problem, which are not restricted to the neighborhood of the
transitions, and which can be used to draw global phase diagrams, in terms of
temperature and either concentration or chemical potential. Also, it is
feasible to extend these calculations beyond the mean-field level, as it has
been done in the annealed case for the analogous model with restricted
orientations \cite{Carmo2}. We use the model solutions to write a Landau-de
Gennes expansion for the free energy in terms of the invariants of the tensor
order parameter, but with model-dependent coefficients. Corresponding
phenomenological expansions have been investigated by a number of authors
\cite{Gramsbergen}\cite{AllenderLonga}\cite{Mukherjee}, and we can use some
asymptotic results to confirm the numerical analysis in the neighborhood of a
Landau multicritical point.

\section{Disordered Maier-Saupe model}

The Maier-Saupe theory of the nematic-isotropic phase transition can be
obtained from a statistical calculation for a fully-connected lattice model
given by the quadrupolar Hamiltonian%
\begin{equation}
\mathcal{H}\left\{  \overrightarrow{n_{i}}\right\}  =-\sum_{1\leq i<j\leq
N}\frac{A}{N}\sum_{\mu,\nu=x,y,z}S_{i}^{\mu\nu}S_{j}^{\mu\nu},
\end{equation}
where the sum is over all pairs of lattice sites, $A/N$ is a scaled
interaction energy, and $\left\{  S_{i}^{\mu\nu}\right\}  $ is a set
orientational (quadrupolar) variables, given by%
\begin{equation}
S_{i}^{\mu\nu}=\frac{1}{2}\left(  3n_{i}^{\mu}n_{i}^{\nu}-\delta_{\mu\nu
}\right)  ,
\end{equation}
where $\overrightarrow{n_{i}}=\left(  n_{i}^{x},n_{i}^{y},n_{i}^{z}\right)  $,
with $\left\vert \overrightarrow{n_{i}}\right\vert =1$, is a local nematic
director. This mean-field Maier-Saupe (MS) model is known to reproduce the
main features of the (weak) first-order transition between uniaxial nematic
and disordered phases.

We mimic the behavior of a binary mixture of oblate and prolate molecules
(discs and cylinders) by introducing an additional set of (shape) degrees of
freedom, $\left\{  \lambda_{i}\right\}  $, with $\lambda_{i}=\pm1$, for
$i=1,...,N$. Given the configurations of orientational and disorder (shape)
degrees of freedom, $\left\{  \overrightarrow{n_{i}}\right\}  $ and $\left\{
\lambda_{i}\right\}  $, the simplest Maier-Saupe Hamiltonian for this mixture
of discs and cylinders is given by%
\begin{equation}
\mathcal{H}\left(  \left\{  \lambda_{i}\right\}  ,\left\{  \overrightarrow
{n_{i}}\right\}  \right)  =-\sum_{1\leq i<j\leq N}\frac{A}{N}\lambda
_{i}\lambda_{j}\sum_{\mu,\nu=x,y,z}S_{i}^{\mu\nu}S_{j}^{\mu\nu},
\end{equation}
which can also be written in the more convenient form%
\begin{equation}
\mathcal{H}\left(  \left\{  \lambda_{i}\right\}  ,\left\{  \overrightarrow
{n_{i}}\right\}  \right)  =-\frac{A}{2N}\sum_{\mu,\nu=x,y,z}\left[  \sum
_{i=1}^{N}\lambda_{i}S_{i}^{\mu\nu}\right]  ^{2},
\end{equation}
where we have discarded irrelevant terms in the $N\rightarrow\infty$ limit.

In a typical problem of a disordered system of solid state physics, the
disordered degrees of freedom are fixed, frozen, while the orientational
degrees of freedom are allowed to thermalize during the experimental times
\cite{Binder}. In this fixed, quenched case, as in amorphous and glassy
materials, disorder variables are not strictly thermodynamic. In the opposite
case, which seems more adequate to describe liquid mixtures, both
orientational and disorder degrees of freedom are allowed to reach
thermodynamic equilibrium during experimental times. The fully thermalized,
annealed case, is then treated according to the standard rules of equilibrium thermodynamics.

In the following paragraphs, we consider quenched and annealed cases
separately. As in the work of Henriques and Henriques \cite{Henriques}, for a
lattice Maier-Saupe model with restricted orientations, we confirm that there
is a biaxial nematic structure in the quenched case. Also, we show that this
biaxial structure becomes thermodynamically unstable in the annealed case,
which agrees with an older Maier-Saupe calculation for a mixture of cylinders
and discs by Palffy-Muhoray and collaborators \cite{Palffy}. We then introduce
the two-temperature formalism \cite{Carmo1}\cite{Nieuwenhuizen} in order to
show that a slight departure from complete thermalization is already
sufficient to produce a stable biaxial nematic phase. The numerical analysis
of the free energy is supplemented, and confirmed, by an analysis of a
Landau-de Gennes expansion in the neighborhood of the Landau multicritical point.

\subsection{Quenched disorder}

Given the set of disorder variables, $\left\{  \lambda_{i}\right\}  $, we
write the canonical partition function%
\begin{equation}
Z\left(  \left\{  \lambda_{i}\right\}  \right)  =%
{\displaystyle\sum\limits_{\left\{  \overrightarrow{n_{i}}\right\}  }}
\exp\left\{  \frac{\beta}{2N}\sum_{\mu,\nu}\left[  \sum_{i=1}^{N}\lambda
_{i}S_{i}^{\mu\nu}\right]  ^{2}\right\}  ,
\end{equation}
where $\beta=A/k_{B}T=1/t$ is the inverse of a (dimensionless) temperature,
$\mu,\nu=1$, $2$, $3$ correspond to the Cartesian directions, and we are
summing over configurations of the local (microscopic) directors. In this
quenched case, $\left\{  \lambda_{i}\right\}  $ is a\ set of independent,
identical, and identically distributed random variables, associated with a
probability distribution%
\begin{equation}
P\left(  \left\{  \lambda_{i}\right\}  \right)  =%
{\displaystyle\prod\limits_{i=1}^{N}}
p\left(  \lambda_{i}\right)  .
\end{equation}

It is convenient to parametrize the local directors by polar coordinates,
\begin{equation}
\overrightarrow{n_{i}}=\left(  \sin\theta_{i}\cos\phi_{i},\sin\theta_{i}%
\sin\phi_{i},\cos\theta_{i}\right)  ,
\end{equation}
with $\Omega_{i}=\left(  \theta_{i},\phi_{i}\right)  $, so that%
\begin{equation}
S_{i}^{\mu\nu}=\frac{3}{2}\left(
\begin{array}
[c]{ccc}%
\sin^{2}\theta_{i}\cos^{2}\phi_{i}-\frac{1}{3} & \sin^{2}\theta_{i}\sin
\phi_{i}\cos\phi_{i} & \sin\theta_{i}\cos\theta_{i}\cos\phi_{i}\\
\sin^{2}\theta_{i}\sin\phi_{i}\cos\phi_{i} & \sin^{2}\theta_{i}\sin^{2}%
\phi_{i}-\frac{1}{3} & \sin\theta_{i}\cos\theta_{i}\sin\phi_{i}\\
\sin\theta_{i}\cos\theta_{i}\cos\phi_{i} & \sin\theta_{i}\cos\theta_{i}%
\sin\phi_{i} & \cos^{2}\theta_{i}-\frac{1}{3}%
\end{array}
\right)  ,\label{suv}%
\end{equation}
and the sum over orientational configurations becomes an integral over solid
angles,%
\begin{equation}
Z\left(  \left\{  \lambda_{i}\right\}  \right)  =\prod_{i}\int d\Omega_{i}%
\exp\left\{  \frac{\beta}{2N}\sum_{\mu,\nu=x,y,z}\left[  \sum_{i=1}^{N}%
\lambda_{i}S^{\mu\nu}\left(  \Omega_{i}\right)  \right]  ^{2}\right\}
.\label{canonical}%
\end{equation}

In the thermodynamic limit ($N\rightarrow\infty$), we have the asymptotic form%
\begin{equation}
Z\left(  \left\{  \lambda_{i}\right\}  \right)  \sim\exp\left[  -\beta
Ng_{N}\left(  \left\{  \lambda_{i}\right\}  \right)  \right]  .
\end{equation}
The resulting (quenched) free energy $g_{q}$ comes from an average of
$g_{N}\left(  \left\{  \lambda_{i}\right\}  \right)  $ over the distribution
of shape variables $P\left(  \left\{  \lambda_{i}\right\}  \right)  $,%
\begin{equation}
g_{q}\sim\left\langle g\left(  \left\{  \lambda_{i}\right\}  \right)
\right\rangle \sim\frac{1}{N}\left\langle \ln Z\left(  \left\{  \lambda
_{i}\right\}  \right)  \right\rangle =\frac{1}{N}\int\left(  \prod_{i}%
d\lambda_{i}\right)  P\left(  \left\{  \lambda_{i}\right\}  \right)  \ln
Z\left(  \left\{  \lambda_{i}\right\}  ,\beta\right)  ,\label{quenched_Gibbs}%
\end{equation}
where the brackets $\left\langle ...\right\rangle $ indicate disorder
averages, and we are taking the limit of large $N$.

The sum over the square terms in equation (\ref{canonical}), can be dealt with
by a set of Gaussian identities. For example, we have%
\[
\exp\left\{  \frac{\beta}{2N}\left[  \sum_{i}\lambda_{i}S^{11}\left(
\Omega_{i}\right)  \right]  ^{2}\right\}  =
\]%
\[
=%
{\displaystyle\int\limits_{-\infty}^{+\infty}}
\frac{dx_{11}}{\sqrt{\pi}}\exp\left\{  -x_{11}^{2}+2\left(  \frac{\beta}%
{2N}\right)  ^{1/2}\left[  \sum_{i=1}^{N}\lambda_{i}S^{11}\left(  \Omega
_{i}\right)  \right]  x_{11}\right\}  =
\]%
\begin{equation}
=\left(  \frac{\beta N}{2\pi}\right)  ^{1/2}%
{\displaystyle\int\limits_{-\infty}^{+\infty}}
dq_{11}\exp\left\{  -\frac{1}{2}N\beta q_{11}^{2}+\sum_{i=1}^{N}\beta
\lambda_{i}S^{11}\left(  \Omega_{i}\right)  q_{11}\right\}  .
\end{equation}
Taking into account the symmetry of the traceless tensor $S^{\mu\nu}$, we
introduce a set of six variables, $q_{11}$, $q_{22}$, $q_{33}$, $q_{12}$,
$q_{13}$, and $q_{23}$, and write the partition function%
\begin{equation}
Z\left(  \left\{  \lambda_{i}\right\}  \right)  =\int\left[  dq\right]
\exp\left\{  -\frac{1}{2}N\beta\sum_{\mu}q_{\mu\mu}^{2}-N\beta\sum_{\mu<\nu
}q_{\mu\nu}^{2}+%
{\displaystyle\sum\limits_{i=1}^{N}}
\ln M_{i}\right\}  ,
\end{equation}
where%
\begin{equation}
\left[  dq\right]  =\left(  \frac{\beta N}{2\pi}\right)  ^{3}dq_{11}%
dq_{22}dq_{33}dq_{12}dq_{13}dq_{23},
\end{equation}
and%
\begin{equation}
M_{i}=M\left(  \lambda_{i},\left\{  q_{\mu\nu}\right\}  \right)  =\int
d\Omega_{i}\exp\left[
{\displaystyle\sum\limits_{\mu\leq\nu}}
\beta\lambda_{i}S^{\mu\nu}\left(  \Omega_{i}\right)  q_{\mu\nu}\right]  .
\end{equation}

In the thermodynamic limit, we resort to Laplace%
\'{}%
s asymptotic method, and invoke the law of large numbers,%
\begin{equation}
\frac{1}{N}%
{\displaystyle\sum\limits_{i=1}^{N}}
\ln M\left(  \lambda_{i},\left\{  q_{\mu\nu}\right\}  \right)  \longrightarrow
\left\langle \ln M\left(  \lambda_{i},\left\{  q_{\mu\nu}\right\}  \right)
\right\rangle =\int d\lambda p\left(  \lambda\right)  \ln M\left(
\lambda,\left\{  q_{\mu\nu}\right\}  \right)  .
\end{equation}
We then have a self-averaged expression for the quenched free energy,%
\begin{equation}
g_{q}=\frac{1}{2}\left(  q_{11}^{2}+q_{22}^{2}+q_{33}^{2}\right)
+q_{12}+q_{13}+q_{23}-\frac{1}{\beta}\int d\lambda p\left(  \lambda\right)
\ln M\left(  \lambda,\left\{  q_{\mu\nu}\right\}  \right)  ,
\end{equation}
where the set of parameters $\left\{  q_{\mu\nu}\right\}  $ come from the
minima of the asymptotic integration,%
\begin{equation}
q_{\delta\delta}=\int\lambda p\left(  \lambda\right)  d\lambda\frac{\int
d\Omega S^{\delta\delta}\left(  \Omega\right)  \exp\left[
{\displaystyle\sum\limits_{\mu\leq\nu}}
\beta\lambda S^{\mu\nu}\left(  \Omega\right)  q_{\mu\nu}\right]  }{\int
d\Omega\exp\left[
{\displaystyle\sum\limits_{\mu\leq\nu}}
\beta\lambda S^{\mu\nu}\left(  \Omega\right)  q_{\mu\nu}\right]  },
\end{equation}
for $\delta=1,2,3$, and
\begin{equation}
q_{\delta\gamma}=\frac{1}{2}\int\lambda p\left(  \lambda\right)  d\lambda
\frac{\int d\Omega S^{\delta\gamma}\left(  \Omega\right)  \exp\left[
{\displaystyle\sum\limits_{\mu\leq\nu}}
\beta\lambda S^{\mu\nu}\left(  \Omega\right)  q_{\mu\nu}\right]  }{\int
d\Omega\exp\left[
{\displaystyle\sum\limits_{\mu\leq\nu}}
\beta\lambda S^{\mu\nu}\left(  \Omega\right)  q_{\mu\nu}\right]  },
\end{equation}
for $\delta<\gamma$. The set of variables $\left\{  q_{\mu\nu}\right\}  $ has
a clear physical interpretation as the mean values of the quadrupole tensor
components. In fact, if we include field terms in the original Hamiltonian, of
the form $h^{\mu\nu}\lambda_{i}S_{i}^{\mu\nu}$, with couplings of local
quadrupoles $\lambda_{i}S_{i}^{\mu\nu}$ to external fields $h^{\mu\nu}$, the
(Gibbs) free energy will depend on these external fields, and the mean
quadrupoles will be given by $q_{\mu\nu}=-\partial g_{q}/\partial h^{\mu\mu}$.

Using the explicit forms of $S^{\mu\nu}\left(  \Omega\right)  $, given by
equation (\ref{suv}), it is straightforward to show that we can choose
$q_{12}=q_{13}=q_{23}=0$, with $q_{11},\,q_{22},\,q_{33}\neq0$. This
self-consistent choice leads to a diagonal mean-quadrupole tensor in a
convenient laboratory frame of reference. We then write the quenched free
energy%
\begin{equation}
g_{q}=\frac{1}{2}\left(  q_{11}^{2}+q_{22}^{2}+q_{33}^{2}\right)  -\frac
{1}{\beta}\int p\left(  \lambda\right)  d\lambda\ln\left\{  \int d\Omega
\exp\left[  \beta\lambda\sum_{\mu=1,2,3}S^{\mu\mu}\left(  \Omega\right)
q_{\mu\mu}\right]  \right\}  ,\label{g_quenched}%
\end{equation}
where%
\begin{equation}
q_{\mu\mu}=\int\lambda p\left(  \lambda\right)  d\lambda\frac{\int d\Omega
S^{\mu\mu}\left(  \Omega\right)  \exp\left[
{\displaystyle\sum\limits_{\mu}}
\beta\lambda S^{\mu\mu}\left(  \Omega\right)  q_{\mu\mu}\right]  }{\int
d\Omega\exp\left[
{\displaystyle\sum\limits_{\mu}}
\beta\lambda S^{\mu\mu}\left(  \Omega\right)  q_{\mu\mu}\right]
}.\label{quu_quenched}%
\end{equation}
Also, we remark that%
\begin{equation}
\sum_{\mu}q_{\mu\mu}=0,
\end{equation}
which confirms the traceless property of the mean-quadrupole tensor.
Therefore, we introduce the standard parametrization%
\begin{equation}
\mathbf{q}=\left(
\begin{array}
[c]{ccc}%
q_{11} & 0 & 0\\
0 & q_{22} & 0\\
0 & 0 & q_{33}%
\end{array}
\right)  =\frac{1}{2}\left(
\begin{array}
[c]{ccc}%
\eta-s & 0 & 0\\
0 & -\eta-s & 0\\
0 & 0 & 2s
\end{array}
\right)  ,\label{standard_param}%
\end{equation}
so that (i) $s\neq0$ and $\eta\neq0$ in a biaxial nematic phase, (ii) $s\neq0$
and $\eta=0$ in a uniaxial nematic phase, and (iii) $s=0$ and $\eta=0$ in the
disordered phase.

The analysis of the quenched free energy depends on the choice of the
distribution $p\left(  \lambda\right)  $. For example, we may choose%
\begin{equation}
p\left(  \lambda\right)  =c\delta\left(  \lambda-1\right)  +\left(
1-c\right)  \delta\left(  \lambda+1\right)  ,\label{double_delta}%
\end{equation}
which represents a sample with a number concentration $c$ of prolate molecules
($\lambda=+1$) and $1-c$ of oblate molecules ($\lambda=-1$), and which is
convenient for comparisons with the annealed situation. If we adopt this form
of $p\left(  \lambda\right)  $, it is straightforward to analyze equations
(\ref{g_quenched}) and (\ref{quu_quenched}), with the standard parametrization
(\ref{standard_param}), and draw the phase diagram of figure (1). We indicate
two uniaxial nematic phases, $N_{+}$ with $s>0$, and $N_{-}$ with $s<0$,
separated by a first-order boundary (heavy dashed line) from the isotropic
phase. The biaxial nematic region is limited by two critical lines that meet
at the Landau multicritical point ($c_{L}=1/2$ and $t_{L}=1/\beta_{L}=3/10$).
In the neighborhood of this Landau point, we confirm these results by the
analysis of an expansion of the free energy in terms of the invariants of the
tensor order parameter.%

\begin{figure}
[ptb]
\begin{center}
\includegraphics[
height=2.6117in,
width=3.7628in
]%
{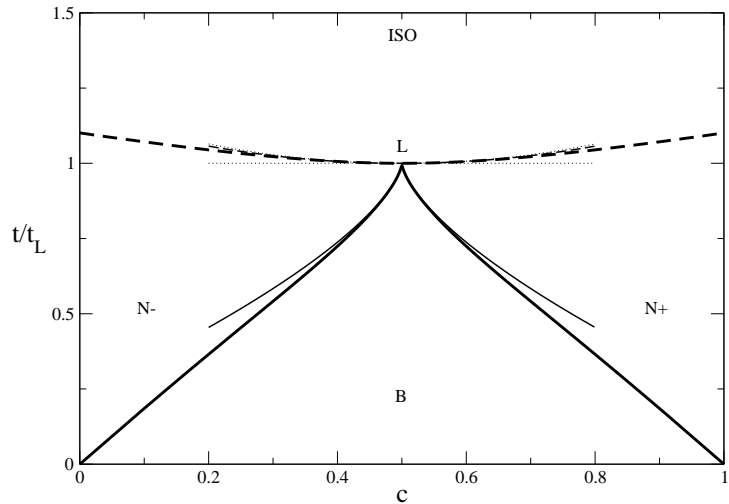}%
\caption{Phase diagram, in terms of the ration between temperature $t/t_{L}$
and the convcentration $c$, for the case of quenched disorder with a
double-delta distribution. We indicate the nematic biaxial ($B$), two uniaxial
nematic phases of opposite symmetry ($N_{+}$ and $N_{-}$) and the Landau
multicritical point ($L$), located at $t_{L}=3/10$ and $c_{L}=0.5$. The heavy
dashed line corresponds to first-order transitions. Heavy solid lines
correspond to continuous transitions (between biaxial an uniaxial nematic
structures). We also indicate asymptotic results coming from an expansion of
the free energy in the neighborhood ot the Landau multicritical point (thin
solid lines). The stability curves for the isotropic and nematic phases are
given by the dotted lines.}%
\end{center}
\end{figure}

\subsection{Annealed disorder}

In the annealed approach, we write the canonical partition function%
\[
Z=\underset{\left\{  \lambda_{i}\right\}  }{%
{\displaystyle\sum}
}^{\prime}\underset{\left\{  \overrightarrow{n}_{i}\right\}  }{\sum}%
\exp\left\{  \frac{\beta}{2N}\sum_{\mu,\nu}\left[  \sum_{i=1}^{N}\lambda
_{i}S_{i}^{\mu\nu}\right]  ^{2}\right\}  ,
\]
where the sum over the shape configurations $\left\{  \lambda_{i}\right\}  $
is restricted by the fixed value of the number density $c$ of prolate
molecules,%
\[%
{\displaystyle\sum\limits_{i=1}^{N}}
\lambda_{i}=N_{+}-N_{-}=N\left(  2c-1\right)  .
\]
It is then convenient to introduce a chemical potential $\mu$ and to change to
a grand canonical ensemble,
\[
\Xi=\underset{\left\{  \lambda_{i}\right\}  }{%
{\displaystyle\sum}
}\underset{\left\{  \overrightarrow{n}_{i}\right\}  }{\sum}\exp\left\{
\frac{1}{2}\beta\mu\left[  N+%
{\displaystyle\sum\limits_{i=1}^{N}}
\lambda_{i}\right]  +\frac{\beta}{2N}\sum_{\mu,\nu}\left[  \sum_{i=1}%
^{N}\lambda_{i}S_{i}^{\mu\nu}\right]  ^{2}\right\}  .
\]

In analogy with the treatment of the quenched case, we use a polar
parametrization for $S^{\mu\nu}$, take advantage of the Gaussian identities to
eliminate the squares, and write the asymptotic ($N\rightarrow\infty$) result%
\begin{equation}
\Xi\sim\exp\left[  -\beta N\phi\right]  ,\label{grand_partition}%
\end{equation}
where $\phi$ is a grand potential per molecule,%
\begin{equation}
\phi=-\frac{\mu}{2}+\frac{1}{2}\left(  q_{11}^{2}+q_{22}^{2}+q_{33}%
^{2}\right)  -\frac{1}{\beta}\ln\zeta\label{GC_function}%
\end{equation}
with%
\[
\zeta=\left[  \exp\left(  \frac{1}{2}\beta\mu\right)  \right]  \int
d\Omega\exp\left[  \beta L\left(  \Omega,\left\{  q_{\mu\mu}\right\}  \right)
\right]  +
\]%
\begin{equation}
+\left[  \exp\left(  -\frac{1}{2}\beta\mu\right)  \right]  \int d\Omega
\exp\left[  -\beta L\left(  \Omega,\left\{  q_{\mu\mu}\right\}  \right)
\right] \label{zeta}%
\end{equation}
and
\begin{equation}
L\left(  \Omega,\left\{  q_{\mu\mu}\right\}  \right)  =S^{11}\left(
\Omega\right)  q_{11}+S^{22}\left(  \Omega\right)  q_{22}+S^{33}\left(
\Omega\right)  q_{33},\label{L}%
\end{equation}
which should be supplemented by the coupled equations for the minimization of
$\phi$ with respect to $q_{11}$, $q_{22}$ and $q_{33}$. Again, we see that the
mean quadrupole tensor is traceless, so that we can use the standard
parametrization of equation (\ref{standard_param}).

The analysis of the free energy shows that the biaxial solution ($s\neq0$,
$\eta\neq0$) is thermodynamically unstable (it is a kind of saddle-point
instead of a minimum of $\phi$). We then draw the phase diagram of figure (2),
in terms of the thermodynamic field variables $t=1/\beta$, dimensionless
temperature, and chemical potential $\mu$. The dashed lines indicate
first-order boundaries between the uniaxial nematic phases $N_{+}$ and $N_{-}%
$, and between the isotropic and each one of the nematic phases. The
multicritical Landau point (at $\mu=0$ and $t_{L}=3/10$) is just a simple
triple point. We can also draw the phase diagram shown in figure (3), in terms
of temperature and concentration, which may be more interesting from the
experimental point of view. The tie lines in the ordered region indicate the
coexistence of two distinct uniaxial nematic phases. Again, we confirm these
results by an analysis of an expansion of the free energy in terms of the
invariants of the tensor order parameter.%

\begin{figure}
[ptb]
\begin{center}
\includegraphics[
height=2.6576in,
width=4.0465in
]%
{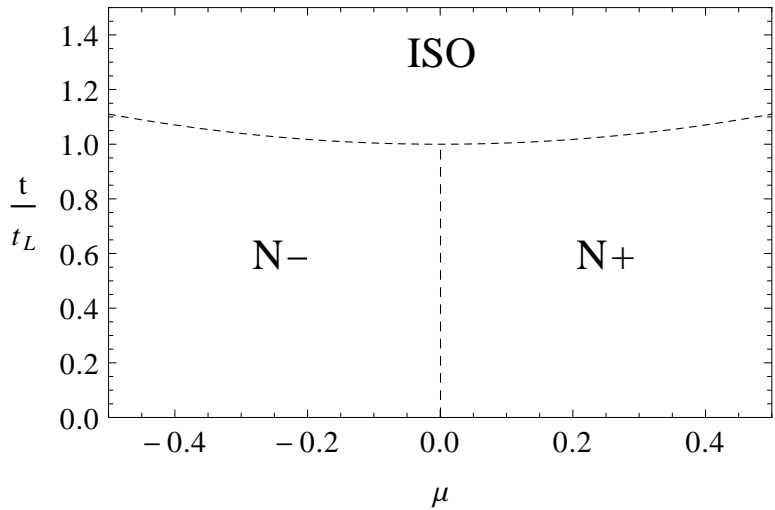}%
\caption{Phase diagram for the annealed case in terms of thermodynamic field
variables (dimensionless temperature $t=1/\beta$ and chemical potential $\mu
$). The triple point is located at $t_{L}=3/10$ and $\mu=0$. Dashed lines
indicate first-order boundaries.}%
\end{center}
\end{figure}
%

\begin{figure}
[ptb]
\begin{center}
\includegraphics[
height=2.6117in,
width=3.7905in
]%
{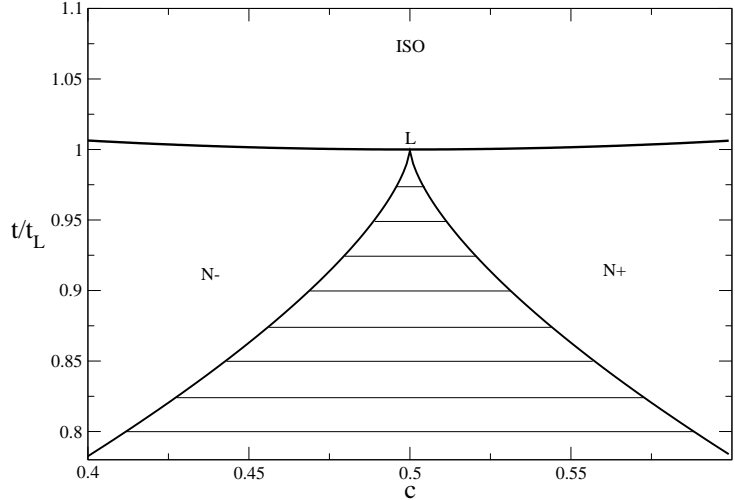}%
\caption{Phase diagram, in terms of temperature and concentration, in the
annealed case. The tie lines indicate the coexistence of two uniaxial nematic
phases. The region of coexistence of the uniaxial nematic and the isotropic
phases is too narrow to be represented in this graph.}%
\end{center}
\end{figure}

\subsection{Two-temperature formalism}

In the two-temperature formalism, we introduce two heat baths, at different
temperatures, associated with the relaxation times of the orientational
(quicker) and disorder (slower) degrees of freedom. We now give a brief
account of this formalism \cite{Nieuwenhuizen}. Given a configuration
$\left\{  \lambda_{i}\right\}  $ of the slower disorder variables, we can
schematically write the probability of occurrence of a configuration $\left\{
\sigma_{i}\right\}  $ of the orientational variables,%
\begin{equation}
P\left(  \left\{  \sigma_{i}\right\}  \left\vert \left\{  \lambda_{i}\right\}
\right.  \right)  =\frac{1}{Z_{\sigma}}\exp\left[  -\beta\mathcal{H}\left(
\left\{  \sigma_{i}\right\}  ,\left\{  \lambda_{i}\right\}  \right)  \right]
,
\end{equation}
where $T=1/\beta$ is the temperature of a heat bath, and
\begin{equation}
Z_{\sigma}=Z_{\sigma}\left(  \left\{  \lambda_{i}\right\}  \right)
=\sum_{\left\{  \sigma_{i}\right\}  }\exp\left[  -\beta\mathcal{H}\left(
\left\{  \sigma_{i}\right\}  ,\left\{  \lambda_{i}\right\}  \right)  \right]
.
\end{equation}
The time evolution of $\lambda_{i}$ is given by a Langevin equation,%
\begin{equation}
\Gamma\frac{\partial\lambda_{i}}{\partial t}=-z\left(  t\right)  \lambda
_{i}-\frac{\partial\mathcal{H}}{\partial\lambda_{i}}+\eta_{i}\left(  t\right)
,
\end{equation}
where $z\left(  t\right)  $ is a multiplier associated with the chemical
potential, and
\begin{equation}
\left\langle \eta_{i}\left(  t\right)  \eta_{j}\left(  t^{\prime}\right)
\right\rangle =2\Gamma T_{\lambda}\delta_{ij}\delta\left(  t%
\acute{}%
-t^{\prime}\right)  ,
\end{equation}
where we have introduced the temperature $T_{\lambda}$ of a second heat bath.
With the assumption of quick and slow time scales, it is reasonable to replace
$\partial\mathcal{H}/\partial\lambda_{i}$ by its average value,%
\begin{equation}
\frac{\partial\mathcal{H}}{\partial\lambda_{i}}\implies\left\langle
\frac{\partial\mathcal{H}}{\partial\lambda_{i}}\right\rangle _{\sigma}%
=\frac{\partial\mathcal{H}_{eff}}{\partial\lambda_{i}},
\end{equation}
where%
\begin{equation}
\mathcal{H}_{eff}=\mathcal{H}_{eff}\left(  \left\{  \lambda_{i}\right\}
\right)  =-k_{B}T\ln%
{\displaystyle\sum\limits_{\left\{  \sigma_{i}\right\}  }}
\exp\left[  -\beta\mathcal{H}\left(  \left\{  \sigma_{i}\right\}  ,\left\{
\lambda_{i}\right\}  \right)  \right]  .
\end{equation}
We then assume that the probability of a configuration $\left\{  \lambda
_{i}\right\}  $ is given by the grand-canonical expression%
\begin{equation}
P\left(  \lambda\right)  =\frac{1}{\Xi\left(  \beta_{\lambda},\beta
,N,\mu\right)  }\exp\left[  \beta_{\lambda}\mu N_{p}-\beta_{\lambda
}\mathcal{H}_{eff}\right]  ,
\end{equation}
where
\begin{equation}
\Xi\left(  \beta_{\lambda},\beta,N,\mu\right)  =\int\left[  d\lambda\right]
\left\{  \sum_{\left\{  \sigma\right\}  }\exp\left[  -\beta\mathcal{H}\left(
\left\{  \sigma_{i}\right\}  ,\left\{  \lambda_{i}\right\}  \right)
+\frac{\beta\mu}{2}\left(
{\textstyle\sum\limits_{i}}
\lambda_{i}+N\right)  \right]  \right\}  ^{\mathbf{n}},
\end{equation}
and the ratio $n=T/T_{\lambda}$ resembles the number of replicas in spin-glass
problems \cite{Binder}.

According to this two-temperature formalism, the orientational degrees of
freedom of the Maier-Saupe model are allowed to reach thermal equilibrium at a
temperature $T$ and the disorder degrees of freedom thermalize at a
temperature $T_{\lambda}\neq T$, with $n=T/T_{\lambda}$. The problem is then
reduced to the calculation of the grand-canonical partition function of $n$
replicas of the original system \cite{Carmo1},%
\[
\Xi_{two}=\underset{\left\{  \lambda_{i,\alpha}\right\}  }{%
{\displaystyle\sum}
}\underset{\left\{  \overrightarrow{n}_{i,\alpha}\right\}  }{\sum}\exp\left\{
%
{\displaystyle\sum\limits_{\alpha=1}^{n}}
\left[  \frac{\beta\mu}{2}\left(  N+%
{\displaystyle\sum\limits_{i=1}^{N}}
\lambda_{i,\alpha}\right)  +\frac{\beta}{2N}\sum_{\mu,\nu}\left(  \sum
_{i=1}^{N}\lambda_{i,\alpha}S_{i,\alpha}^{\mu\nu}\right)  ^{2}\right]
\right\}  .
\]
In the thermodynamic limit, we write%
\begin{equation}
\Xi_{two}\sim\exp\left[  -\beta N\phi_{two}\right]
,\label{grand_canonical_two}%
\end{equation}
where%
\begin{equation}
\phi_{two}=-\frac{1}{2}\mu n+\frac{1}{2}n\left(  q_{11}^{2}+q_{22}^{2}%
+q_{33}^{2}\right)  -\frac{1}{\beta}\ln\zeta_{two}\label{GC_potential_two}%
\end{equation}
with%
\[
\zeta_{two}=\left\{  \left[  \exp\left(  \frac{1}{2}\beta\mu\right)  \right]
\int d\Omega\exp\left[  \beta L\left(  \Omega,\left\{  q_{\mu\mu}\right\}
\right)  \right]  \right\}  ^{n}+
\]%
\begin{equation}
+\left\{  \left[  \exp\left(  -\frac{1}{2}\beta\mu\right)  \right]  \int
d\Omega\exp\left[  -\beta L\left(  \Omega,\left\{  q_{\mu\mu}\right\}
\right)  \right]  \right\}  ^{n}.\label{zeta_two}%
\end{equation}
The minimization of the grand potential $\phi_{two}$ leads to the equilibrium
values $q_{11}$, $q_{22}$, and $q_{33}=-q_{11}-q_{22}$. Note that we regain
the annealed case for $n=1$, and that the role of this parameter $n$ will
become clear in the next Section.

\section{Connections with the Landau-de Gennes expansion}

We have already chosen a standard order parameter, given by the traceless
diagonal tensor (\ref{standard_param}). We then introduce the second and
third-order invariants,%
\begin{equation}
I_{2}=Tr\left[  \mathbf{q}^{2}\right]  =q_{11}^{2}+q_{22}^{2}+q_{33}^{2}%
=\frac{1}{2}\left(  3s^{2}+\eta^{2}\right) \label{I2}%
\end{equation}
and
\begin{equation}
I_{3}=Tr\left[  \mathbf{q}^{3}\right]  =q_{11}^{3}+q_{22}^{3}+q_{33}^{3}%
=\frac{3}{4}s\left(  s^{2}-\eta^{2}\right)  ,\label{I3}%
\end{equation}
in terms of which it is usual to write the phenomenological Landau-de Gennes
expansion for the free energy in the neighborhood of a transition,%
\begin{equation}
f=f_{0}+\frac{A}{2}I_{2}+\frac{B}{3}I_{3}+\frac{C}{4}\left(  I_{2}\right)
^{2}+\frac{D}{5}I_{2}I_{3}+\frac{E}{6}\left(  I_{3}\right)  ^{2}%
+\frac{E^{^{\prime}}}{6}\left(  I_{2}\right)  ^{3}+...\label{Landau_general}%
\end{equation}
The Landau multicritical point is given by $A=B=0$. The stability conditions
of the ordered phases in the neighborhood of the Landau point are discussed in
terms of the signs of the coefficients of the higher-order terms. It has been
shown that $E>0$ is a necessary condition for the stability of a biaxial
nematic phase in the vicinity of the Landau multicritical point
\cite{Gramsbergen}\cite{AllenderLonga}.

In the present case, we remark that it is more convenient to adopt an
alternative parametrization, in terms of two new variables $r$ and $\psi$,
such that
\begin{equation}
q_{11}=-\frac{r}{2}\left(  \cos\psi+\sqrt{3}\sin\psi\right)  ,\label{q1_Alben}%
\end{equation}%
\begin{equation}
q_{22}=-\frac{r}{2}\left(  \cos\psi-\sqrt{3}\sin\psi\right)  ,\label{q2_Alben}%
\end{equation}
and
\begin{equation}
q_{33}=r\cos\psi.\label{q3_Alben}%
\end{equation}
We then have $q_{11}+q_{22}+q_{33}=0$,
\begin{equation}
I_{2}=\frac{3}{2}r^{2}\label{I2r}%
\end{equation}
and
\begin{equation}
I_{3}=\frac{3}{4}r^{3}\cos\left(  3\psi\right)  .\label{I3rpsi}%
\end{equation}

\subsection{Quenched disorder}

In the neighborhood of the Landau multicritical point, the quenched free
energy, given by equation (\ref{g_quenched}), leads to the expansion%
\[
g_{q}=g_{0}+\frac{1}{2}\left(  1-\frac{3\beta\overline{\lambda^{2}}}%
{10}\right)  I_{2}-\frac{10\overline{\lambda^{3}}}{21\left(  \overline
{\lambda^{2}}\right)  ^{2}}I_{3}+
\]%
\begin{equation}
+\frac{5\overline{\lambda^{4}}}{42\left(  \overline{\lambda^{2}}\right)  ^{3}%
}I_{2}^{2}+\frac{100\overline{\lambda^{5}}}{231\left(  \overline{\lambda^{2}%
}\right)  ^{2}}I_{2}I_{3}+\frac{2000\overline{\lambda^{6}}}{7007\left(
\overline{\lambda^{2}}\right)  ^{5}}I_{3}^{2}-\frac{1450\overline{\lambda^{6}%
}}{27027\left(  \overline{\lambda^{2}}\right)  ^{5}}I_{2}^{3}\label{Gq_Landau}%
\end{equation}
where $g_{0}$ is the free energy of the isotropic phase,
\begin{equation}
\overline{\lambda^{k}}=\int\lambda^{k}p\left(  \lambda\right)  d\lambda,
\end{equation}
and we have kept terms up to fifth order only. The Landau multicritical point
is given by $t=3\overline{\lambda^{2}}/10$ and $\overline{\lambda^{3}}=0$ (for
an arbitrary distribution of shapes).

Let us choose the double-delta distribution, given by equation
(\ref{double_delta}), which is particularly adequate for a comparison with the
annealed case. The expansion of the free energy in the neighborhood of the
Landau multicritical point ($\beta=\beta_{L}=10/3$ and $c=c_{L}=1/2$) is given
by%
\[
g_{q}=-\frac{\ln\left(  4\pi\right)  }{\beta}+\frac{1}{2}\left(
1-\frac{3\beta}{10}\right)  I_{2}-\frac{10}{21}\left(  2c-1\right)  I_{3}+
\]

\begin{equation}
+\frac{5}{42}I_{2}^{2}+\frac{2000}{7007}I_{3}^{2}-\frac{1450}{27027}I_{2}%
^{3}.\label{Gq_Landau_c}%
\end{equation}
The positive sign of the coefficient of $I_{3}^{2}$ indicates that biaxial
nematic phase is stable in the neighborhood of this Landau point. Sufficiently
close to the Landau point, we can show that the first-order transition between
the uniaxial nematic and the disordered phase is asymptotically given by
\begin{equation}
\frac{t}{t_{L}}=1+\frac{10}{63}\left(  2c-1\right)  ^{2}%
,\label{coexistence_quenched}%
\end{equation}
in agreement with numerical calculations (see figure 1). Also, we show that
the critical lines separating the biaxial and the two uniaxial nematic phases
are given by
\begin{equation}
\frac{t}{t_{L}}=1-\frac{10}{21}\left(  6\right)  ^{1/3}\left(  \frac
{1001}{1200}\right)  ^{2/3}\left(  2c-1\right)  ^{2/3}%
,\label{critical_quenched}%
\end{equation}
which also agrees with numerical calculations close to the Landau point (see
figure 1). This phase diagram, with a stable biaxial nematic phase, is in
qualitative agreement with previous results for a Maier-Saupe model with
restricted orientations \cite{Henriques}\cite{Carmo1}. Using the notation of
the phenomenological Landau-de Gennes free energy, we remark that $C>0$ and
$D=0$, as in the work of Allender and Longa \cite{AllenderLonga}. However, new
topologies may arise if we consider other forms of the distribution $p\left(
\lambda\right)  $.

\subsection{Annealed disorder}

In the annealed case, we use the grand potential $\phi$, given by equation
(\ref{GC_function}), to locate the Landau multicritical point ($\beta
_{L}=10/3$ and $\mu_{L}=0$), and write the expansion%
\[
\phi=\phi_{0}+\frac{1}{2}\left(  1-\frac{3\beta}{10}\right)  I_{2}-\frac
{50}{63}\mu I_{3}+
\]%
\begin{equation}
+\frac{5}{42}I_{2}^{2}-\frac{2500}{27027}I_{3}^{2}-\frac{1450}{27027}I_{2}%
^{3},\label{GC_Landau}%
\end{equation}
where
\begin{equation}
\phi_{0}=-\frac{1}{2}\mu-\frac{1}{\beta}\ln\left[  8\pi\cosh\left(  \frac
{1}{2}\beta\mu\right)  \right] \label{phi_zero_GC}%
\end{equation}
is the grand potential of the isotropic phase. The negative coefficient of
$I_{3}^{2}$ shows that there is no stable biaxial nematic phase in the
neighborhood of the Landau multicritical point. The line at $\mu=0$, below the
temperature of the Landau point, is a first-order boundary between two
distinct uniaxial nematic phase (with $s>0$ and $s>0$). We show that the
first-order lines separating the uniaxial nematic from the isotropic phase are
given by the asymptotic expression
\begin{equation}
\frac{t}{t_{L}}=1+\frac{250}{567}\mu^{2},\label{coexistence_annealed}%
\end{equation}
in the immediate vicinity of the Landau point. The phase diagram in figure 2
is in agreement with these asymptotic results.

We can also calculate some asymptotic expressions in terms of temperature and
concentration, which are more convenient variables from the experimental point
of view. For example, the region of coexistence of the uniaxial nematic phases
in figure 3 is limited by the asymptotic border
\begin{equation}
\frac{t}{t_{L}}=1-\frac{5}{7}\left(  \frac{42}{25}\right)  ^{2/3}\left(
c-\frac{1}{2}\right)  ^{2/3}.\label{coexistence_figure3}%
\end{equation}
The asymptotic form of the first-order border between the uniaxial nematic and
the isotropic phases is given by%
\begin{equation}
\frac{t}{t_{L}}=1+\frac{40}{63}\left(  c-\frac{1}{2}\right)  ^{2}%
,\label{n-iso_annealed_concentration}%
\end{equation}
in full agreement with numerical calculations. Also, it should be remarked
that the phase diagram in figure 3 is in qualitative agreement with previous
calculations for a Maier-Saupe model with restricted orientations
\cite{Carmo1}.

\subsection{Two-temperature formalism}

The same sort of calculations can be carried out for the grand potential in
the two-temperature formalism. From equation (\ref{GC_potential_two}), in the
immediate neighborhood of the Landau point, we have the expansion%
\[
\phi_{two}=\phi_{0,two}+\frac{1}{2}\left(  1-\frac{3\beta}{10}\right)
nI_{2}-\frac{50}{63}\frac{\mu}{n}I_{3}+
\]%
\begin{equation}
\frac{5}{42}\frac{1}{n^{2}}I_{2}^{2}+\frac{2500}{27027}\frac{1}{n^{4}}\left(
\frac{108}{35}-\frac{143n}{35}\right)  I_{3}^{2}-\frac{1450}{27027}\frac
{1}{n^{4}}I_{2}^{3},\label{GC_two_Landau}%
\end{equation}
where $\phi_{0,two}$ is the grand potential of the isotropic phase, and we
keep terms up to fifth order. Of course, we recover the expansion for the
annealed case with $n=T/T_{\lambda}=1$. The Landau multicritical point is
still located at $\beta_{L}=10/3$ and $\mu_{L}=0$, but the sign of the
coefficient of $I_{3}^{2}$ depends on the parameter $n$. Indeed, there will be
a stable biaxial nematic phase for%
\begin{equation}
n<\frac{108}{143}\approx\frac{2}{3},
\end{equation}
which indicates that a slight departure from complete annealing ($n=1$) is
already enough to give rise to a stable biaxial structure. Comparisons with
previous results for a Maier-Saupe model with restricted orientations, $n<0.9
$, show that a somewhat larger difference between the temperatures is needed
to stabilize the biaxial phase in the presence of additional direction fluctuations.

\section{Conclusions}

We have carried out exact statistical-mechanics calculations for a Maier-Saupe
lattice model with the inclusion of extra disorder degrees of freedom to mimic
a mixture of discs and cylinders. The closed-form solutions can be written as
a Landau-de Gennes expansion for the free energy in the neighborhood of the
transition, with explicit forms of model-dependent coefficients, which allows
the use of several results from the literature. The stability of a biaxial
nematic structure depends on the treatment of the disorder degrees of freedom.
For quenched disorder, with a typical double-delta distribution of discs and
cylinder, we obtain a global phase diagram, in terms of temperature and
concentration, with a Landau multicritical point, a biaxial and two uniaxial
nematic phases. If the disorder degrees of freedom are allowed to reach
thermal equilibrium, we show that the biaxial structure becomes unstable. We
then assume that orientation and disorder degrees of freedom are coupled to
different heat reservoirs, with two different temperatures. In the
two-temperature context, we show that a small temperature difference, which is
equivalent to a small departure from thermalization, is already enough to
stabilize a biaxial nematic structure. These results explain some
disagreements in the earlier literature. Also, they qualitatively agree with a
previous calculation for a similar Maier-Saupe model with restricted
directional orientations.\bigskip

\textbf{Acknowledgments}

We thank Eduardo do Carmo and Danilo Liarte for helpful suggestions and
comments. This work has been supported by grants from FAPESP and CNPq.

\end{document}